# IDSAC – IUCAA Digital Sampler Array Controller

Sabyasachi Chattopadhyay*[a], Pravin Chordia[a], A. N. Ramaprakash[a], Mahesh P. Burse[a], Bhushan Joshi[a], Kalpesh Chillal[a]

[a]Inter-University Centre for Astronomy and Astrophysics, Pune, India

## ABSTRACT

In order to run the large format detector arrays and mosaics that are required by most astronomical instruments, readout electronic controllers are required which can process multiple CCD outputs simultaneously at high speeds and low noise levels. These CCD controllers need to be modular and configurable, should be able to run multiple detector types to cater to a wide variety of requirements. IUCAA Digital Sampler Array Controller (IDSAC), is a generic CCD Controller based on a fully scalable architecture which is adequately flexible and powerful enough to control a wide variety of detectors used in ground based astronomy. The controller has a modular backplane architecture that consists of Single Board Controller Cards (SBCs) and can control up to 5 CCDs (mosaic or independent). Each Single Board Controller (SBC) has all the resources to a run Single large format CCD having up to four outputs. All SBCs are identical and are easily interchangeable without needing any reconfiguration. A four channel video processor on each SBC can process up to four output CCDs with or without dummy outputs at 0.5 Megapixels/Sec/Channel with 16 bit resolution. Each SBC has a USB 2.0 interface which can be connected to a host computer via optional USB to Fibre converters. The SBC uses a reconfigurable hardware (FPGA) as a Master Controller. IDSAC offers Digital Correlated Double Sampling (DCDS) to eliminate thermal kTC noise. CDS performed in Digital domain (DCDS) has several advantages over its analog counterpart, such as - less electronics, faster readout and easier post processing. It is also flexible with sampling rate and pixel throughput while maintaining the core circuit topology intact. Noise characterization of the IDSAC CDS signal chain has been performed by analytical modelling and practical measurements. Various types of noise such as white, pink, power supply, bias etc. has been considered while creating an analytical noise model tool to predict noise of a controller system like IDSAC. Several tests are performed to measure the actual noise of IDSAC. The theoretical calculation matches very well with practical measurements within 10% accuracy.

## 1. INTRODUCTION

The scientific applications in the field of astronomy increasingly require finer image resolutions. This has evolved as a demand for larger focal plane arrays. As a result, traditional CCD readout techniques which employ analog Correlated Double Sampling (CDS) become large and cumbersome. CDS in digital domain comes with much higher throughput with far less noise and is relatively compact. It is also flexible with sampling rate and pixel throughput while maintaining the core circuit topology. Although DCDS approaches have made good progress through work done by Wu et al[1], Gach et al[2], Clapp[3], Obroslak et al[4], etc., the characterization of DCDS induced noises with a thorough support of analytical calculations is yet to be done. We have carried out a full characterization of the Digital CDS system implemented in IDSAC and compared it with actual measurements to achieve an end to end understanding.

In Section 2 we will present the design of IDSAC and its functional aspects. We will discuss an analytical noise model of DCDS and how it is implemented in IDSAC in Sections 3 & 4. Section 5 primarily discusses various tests which were performed to characterize IDSAC. Although we have tried to optimize our system as the controller for the eight spectrographs of the Devasthal Optical Telescope Integral Field Spectrograph (DOTIFS) instrument, our approach was to achieve optimal noise performance at the highest possible CCD throughput.

## 2. IDSAC DESIGN AND FUNCTIONALITY

The IDSAC system has been developed at IUCAA for controlling a wide variety of astronomical detectors like CCD, EMCCD, CMOS, and IR etc. IDSAC is a multi-controller, multi-interface, scalable system, capable of controlling all the above mentioned detector types. Specific configurations of the controller can be chosen depending on the type and number of detectors to be controlled. In this section we will describe the various hardware and firmware subsystems of IDSAC in brief and their functionality. We will also touch upon the control software that has been developed for the entire characterization process described in Section 5.

**2.1 System Level Architecture**

Fig. 1 shows the system level representation of the IDSAC controller. This IDSAC version has a backplane Card which can accommodate up to 5 SBCs. It can easily be expanded to handle more SBCs, if required. Each SBC has a USB 2.0 interface which is connected to a separate host computer via optional USB to Fiber convertors. All the SBCs are interchangeable without any reconfiguration.

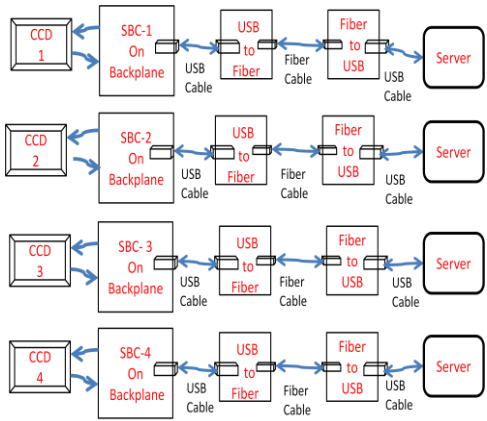
Figure 1: IDSAC Controller

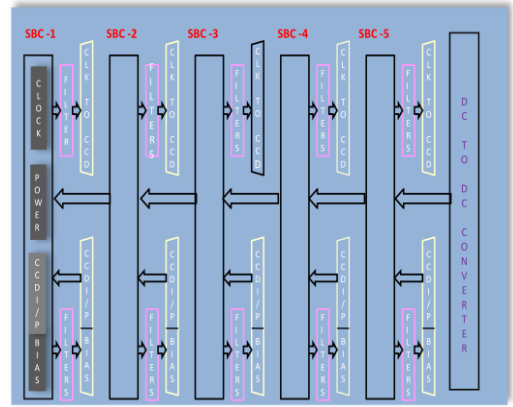
Figure 2: Backplane Block Diagram

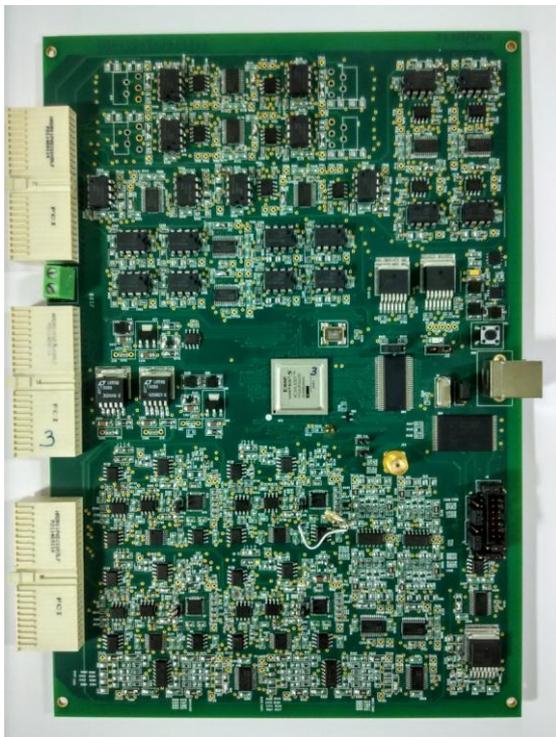
Figure 3: Image of an IDSAC SBC.

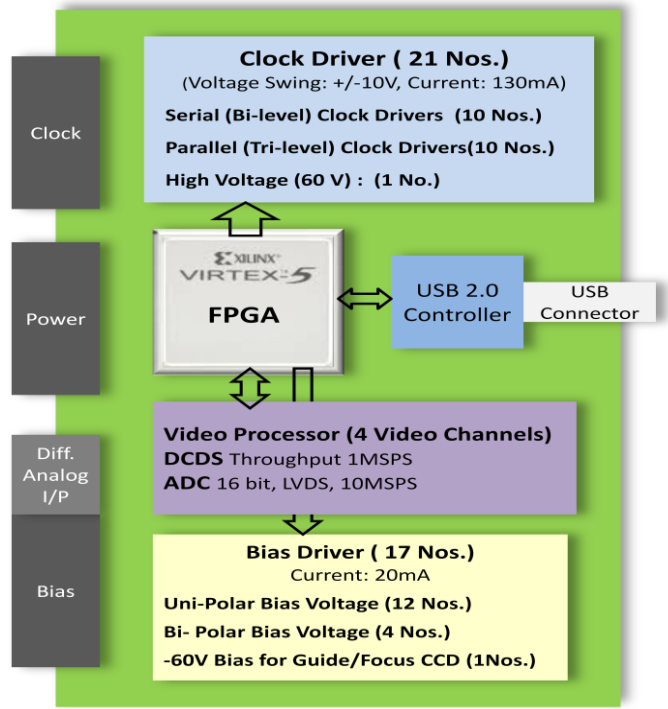
Figure 4: SBC Block Diagram

**2.2 Single Board Controller**

Each of the single board controllers (SBCs) has all the resources required to run a CCD. The block diagram shown in Fig 4 also indicates relative placement of various modules of controller on the board. Based on functionality and placement, the card can be divided in four main subsections.

- **Clock Driver:** This section has 21 high current Clock Drivers, ten of which are Bi-level Serial Clock drivers and ten are Tri-level Parallel Clocks. One additional high voltage clock is provided which can be programmed for operation up to 60V for electron multiplied CCDs and deep depletion CCDs.
    - **Serial or Bi-Level Clock Driver:** Serial clock drivers generate 10 clocks involved in pixel timing: serial register clocks, Summing Well, and Reset Gate. The analog switch, which selects between two bipolar DACs, is followed by a buffer which provides the drive current. The buffer has required gain to enable a clock swing in excess of the DAC range as required by CCD. A control signal for the Analog Switch is generated by an FPGA as a part of timing waveform generation.
    - **Parallel or Tri-Level Clock Driver:** Parallel clock drivers generate 10 clocks involved in line transfer: Image Area Clocks and Transfer gate. Due to the slow nature of these clocks, parallel clock states are changed by updating DAC values. This scheme reduces the number of DACs per clock by half and eliminates the need for Analog switches. It uses a DAC in which the values can be preloaded but they will not appear at the DAC outputs till a strobe signal called LDAC is applied. Upon activating LDAC pin all the DAC outputs gets updated simultaneously. This approach also supports "Slow Tri-level Clocking" which increases Charge transfer efficiency.
- **Master Controller/Digital Section:** Xilinx make Virtex-5 series FPGA is used as an embedded controller for generating readout clocks, reading four 16 bit serial ADCs and programming DACs for bias voltages. USB 2.0 port has been provided as the PC interface for host communication. Algorithm for DCDS is also implemented in the same FPGA having a throughput of 0.5megapixels per second per channel for all four channels.
- **Video Processor/ Digital Correlated Double Sampler (DCDS):** The Video Processor uses Digital Correlated double sampling technique for processing CCD signal. It uses a unity gain differential ADC driver followed by a 16 bit 10MSPS ADC. An output from ADC $V_{CM}$ nulls the common mode signal. Input to Video processor is from a Preamplifier while an analog switch is used for resetting the anti- aliasing filter.
- **Bias Driver:** A generic CCD controller requires 16 output biases, out of which 12 biases can swing from 0 to 30 V and remaining 4 can swing within +/- 14V. An extra negative high voltage output is needed for the backside contact of the delta doped guider/focus CCDs. Substantial filtering has been added at the output of Biases to achieve good noise performance. The Bias drivers are frequency compensated as it drives the big capacitive load filters.

**2.3 Backplane Architecture**

The backplane card provides power supply to all the 5 SBCs which is shown in fig. 2. Except the USB Interface to PC, all the controllers interconnect via this board. It has signal conditioning circuits for all the Clocks and Biases connected to the CCD and Power supply input to the SBCs. Customization of SBCs if any, is achieved through jumper settings on the backplane card.

**2.4 Firmware and Flexibility**

The host software runs on a Linux PC and users can configure the controller on the fly for the current detector system and demand arbitrary ROI readouts along with over scan and dark pixels. To generate ROI readout clocks, IDSAC needs to know some basic information related to the detector's characteristics and intended use. All this information is part of the detector data sheet and has to be captured and passed on to IDSAC in a specified format. This is a onetime process for a given detector. Users will be able to give information through a set of values in different parameter files. IDSAC host software will gather this information from parameter files and send it to the FPGA by generating and sending the required low level commands.

## 3. DCDS ANALYTICAL NOISE MODEL

In this section we will discuss the implementation and an analytical noise model of the IDSAC DCDS signal chain. Any CDS signal chain carries signal from the output of the detector to the input of the Analog to Digital Converter (ADC) which digitizes the analog voltage to binary data.

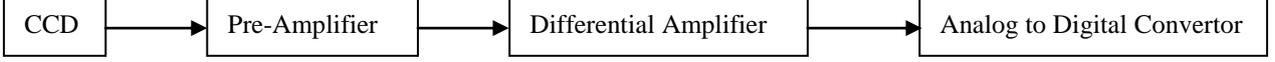

Figure 5: Simplified DCDS (bottom) signal chain.

A simplified DCDS signal chain primarily contains a pre-amplifier block followed by an ADC driver before the ADC itself. For every block (e.g. Preamplifier, ADC driver etc), the noise from each source gets added at that block's output. For a signal chain, which consists of several blocks, noise power gets transmitted through several blocks to the output. The input referred noise is the output referred noise divided by power gain of the signal chain.

Let us consider a generic signal chain that has n electronic blocks. Let $N_1, N_2,..., N_n$ be the input referred noise powers of blocks 1, 2,..., n and $G_1, G_2,..., G_n$ their power gains respectively. If the total noise power at the output is $N_{op}$ then,

$$N_{op} = [N_1 G_1 G_2 \ldots G_n] + [N_2 G_2 \ldots G_n] + \cdots + [N_n G_n]. \tag{1}$$

Now if $N_{ip}$ is the input referred noise power of the entire signal chain, we can write that,

$$N_{op} = [N_{ip} G_1 G_2 \ldots G_n]. \tag{2}$$

From equation 1 and 2 we can say that,

$$N_{ip} = N'_1 B_1 + \frac{N'_2 B_2}{G_1} + \cdots + \frac{N'_n B_n}{G_1 G_2 \ldots G_{n-1}}, \tag{3}$$

where $N'_1, N'_2,..., N'_n$ are noise power spectral densities and $B_1, B_2,..., B_n$ are bandwidths of blocks 1, 2,..., n respectively. As is well known and also seen from eq. 3, one can keep the overall input referred noise of the signal chain low by keeping a high gain low noise block at the input of the signal chain.

## 4. NOISE SOURCES

In the previous section we have considered the cumulative noise power for each block. In reality, each of this noise power has different noise components coming from different origins. We have explored the effects various noise sources such as White noise, Pink noise and Power Supply noise in detail. We have also taken other noise sources such as clock jitter and bias level noise into consideration only to find out that effect of these noises have negligible effect on the overall noise performance of the CDS signal chain. In this section we will discuss the contribution of different noise sources separately.

### 4.1 White Noise

For a resistor of resistance R ohm, temperature T Kelvin, operated at bandwidth B Hz, the white noise power will be $N_{WN} = 4 * k * T * R * B$, where k is Boltzmann constant. To keep it simple, one can replace a "noisy" resistor with a "noiseless/ideal" resistor and a voltage source of amplitude $\sqrt{N_{WN}}$ in series. In Fig. 6 the noise model of an OpAmp circuit is shown in inverting configuration. The total output noise power has contribution from all the noise sources.

For this OpAmp block, output referred RMS noise voltage will be,

$$E_{WN} = \left[ \int \left( 4kR_1 \left(\frac{R_1}{R_2}\right)^2 + 4kTR_2 + 4kTR_3 \left(\frac{R_1+R_2}{R_1}\right)^2 + e_n^2 \left(\frac{R_1+R_2}{R_1}\right)^2 + i_{np}^2 \left(\frac{R_1+R_2}{R_1}\right)^2 + i_{nn}^2 R_2^2 \right) df \right]^{\frac{1}{2}}. \tag{4}$$

The bandwidth for the integration is set by the bandwidth of the OpAmp blocks which is either defined by the capacitor resistor circuit implemented in its feedback path or by the OpAmp itself whichever is lower. Although we can calculate noise voltage of an OpAmp block using this method, a block consisting of an ADC needs a different treatment. To measure the ADC noise at the output it is required to know the Signal to Noise Ratio (SNR - given at a specific bandwidth) and Full Scale input power which are mentioned in the data-sheet. So the input referred noise power of an ADC block will be signal power (measured as power of a sine wave of full scale peak to peak amplitude applied as a signal at the input) divided by the SNR.

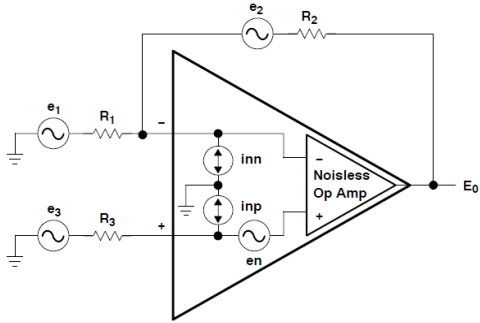

Figure 6: OpAmp Circuit Noise Sources.

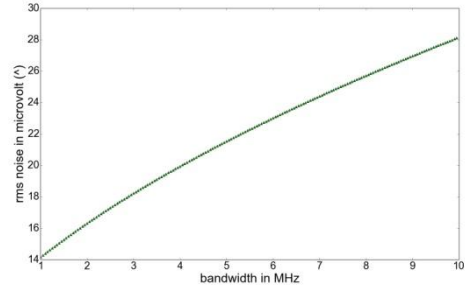

Figure 7: Analytical Noise Prediction for DCDS for 0.5 MPPS CCD throughput and 10 MSPS ADC sampling rate. The number of samples is fixed at 3 for each sampling interval.

### 4.2 Pink Noise

Pink noise PSD varies in inverse proportion to the frequency, which means pink noise dominates at lower frequencies. We have used the noise corner frequency ($f_c$) to calculate PSD of Pink noise. At frequency $f = f_c$, white noise PSD have the same value as that of the pink noise PSD. So the pink noise PSD will be, $N_{pn} = N_{wn} f_c \frac{1}{f}$. To obtain the noise power of pink noise one has to integrate $N_{pn}$ over a frequency range. The lower cut-off frequency of the AC coupler (which acts as a High Pass Filter) at the input of the signal chain determines the $f_L$, while the upper cut-off is $f_c$. Since the contribution of the pink noise will at both ends follow the slow roll off of the filter transfer function, we took $f_L/5$ and $5*f_c$ to be the boundary frequencies. So the RMS noise voltage contribution from the pink noise will be, $E_{PN} = \sqrt{N_{wn} f_c \ln\left(\frac{25 f_c}{f_L}\right)}$. The typical RMS pink noise voltage of a Pre-Amplifier block used for DCDS signal chain in IDSAC is less than 1 µV which is negligible compared to white noise. Thus we can safely neglect the effect of pink noise in those systems where bandwidth is much higher than the noise corner frequency $f_c$.

### 4.3 Power Supply Noise

High frequency noise components present in the power supply can also emerge at the output of the OpAmp. PSRR of an OpAmp provides the insulation of such noise from appearing at OpAmp output. For an OpAmp of PSRR 100 dB, to get a contribution of 1% of output referred noise the RMS noise voltage of a power supply should be less than 30 mV. This noise value is quite high compared to the noise level found in the power supply which is used for the electronic circuits. So we can safely say that the noise contributed from power supply is negligible as well.

### 4.4 Analytical Noise Prediction

To use this analytical noise model of DCDS signal we have also developed a python tool that will analytically predict the noise of a signal chain provided all the components and working conditions are known. This tool not only can provide noise of a particular combination of bandwidth, number of samples in each sampling interval and CCD throughput but also produces plot of noise vs above quantities. Analytical noise prediction for DCDS against varying bandwidth at 0.5MPPS CCD throughput and 3 samples per interval is shown in Fig. 7.

# 5. CHARACTERIZATION OF IDSAC

**5.1 Setup:**

The primary objective of characterization is to run DCDS at the highest possible CCD throughput as well as to optimize bandwidth and number of samples. The system is going to be used to read out a 2kx4k CCD with 15 μm pixels for the Devasthal Optical Telescope Integral Field Spectrograph (DOTIFS). For the characterization we have used a grade 5 e2v CCD42-40 mounted in a cryogenic test Dewar. This CCD has 150,000 electrons full well capacity and an amplifier response of 4.5 μV per electron. We have tested the system at 500 KPPS and 350 KPPS CCD throughputs. A typical test setup is shown in fig. 8.

**5.2 Hardware Characterization:**

Initially the following tests were performed without connecting a CCD to IDSAC. The Analog Chain of the SBC was used for data acquisition.

- To measure the ADC noise we have analyzed image taken by shorting both inputs of the ADC to ground and measured the standard deviation in resultant fits image. It is found that for a CDS throughput of 500 KPPS and 3 samples per sampling interval, the RMS error is 0.58 counts.
- To search for any missing counts in ADC output we have plotted a histogram of pixel values of an exposure in which a slow 10 Hz triangle wave is fed to the input of the signal chain. The amplitude of the triangle wave is selected to cover the entire ADC range. We have not found any missing count.
- Noise of entire analog chain is measured from fits images by grounding both input of signal chain (preamplifier input). The RMS noise is found to be 2.1 counts for a CDS throughput of 500 KPPS and 3 samples in each sampling interval.
- Each analog bias has been tested for multiple bias levels covering full range and with appropriate RC loads.
    - To measure the bias voltage noise we have collected 4M data points for each voltage level at a conversion speed of 1MHz and plotted histograms to obtain the width of the Gaussian profile which is found to be <1mV as expected.
    - Similar tests were performed for the rails of all Clocks.
    - Rise and fall time of Clock drivers were checked using dummy capacitive load simulating present day big CCDs.
- Next, tests were performed after connecting a grade 5 e2v CCD42-40 to two of the inputs of the CDS signal chain. We have performed noise measurement of the entire signal chain for different number of digital samples at each sampling level with 350 KPPS per channel CCD throughput. We have also tested the role of bandwidth by changing the system bandwidth while fixing the CCD throughput at 500 KPPS per channel. The results are tabulated in Table 1 and 2.

Table 1: Characterization of DCDS signal chain at 350 KPPS readout rate. The bandwidth used is 3 MHz.

| Number of Samples in each sampling interval | 3 Samples | 4 Samples | 5 Samples |
|---|---|---|---|
| Noise of Signal chain and CCD in $e^-$ | 7.5 | 7.1 | 6.7 |
| Noise of Signal Chain in $e^-$ | 5.4 | 4.7 | 4.1 |

Table 2: Effect of bandwidth in DCDS rms noise. The test is performed at 500 KPPS readout rate for which the CCD noise is 6 electrons. There are 3 samples in each sampling interval.

| Bandwidth in MHz | 7MHz | 4.8MHz | 3MHz | 1.9MHz |
|---|---|---|---|---|
| Noise of Signal Chain and CCD in $e^-$ | 7.7 | 7.4 | 7.09 | 7.2 |
| Noise of Signal Chain in $e^-$ | 4.8 | 4.4 | 3.8 | 4 |

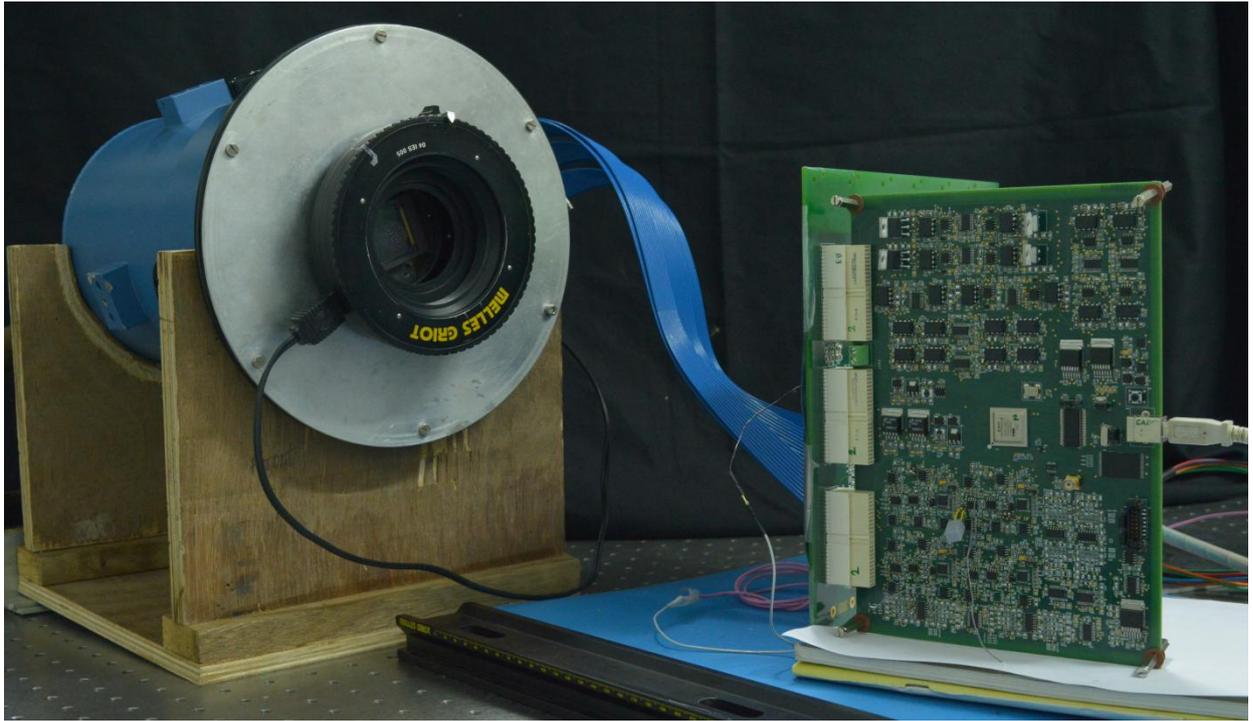

Figure 8: Test setup for IDSAC Characterization

## 6. SUMMARY

We have found that noise of the signal chain is 5.4 electrons which is highest measured noise of the analog signal chain. So each video processor channel provides better noise performance than the CCD as the CCD noise at 500 KPPS is 6 electrons. All the noise measurements are done simultaneously for both the outputs of the CCD using 2 different video processor channels. The noise performances provided in Tables 1 and 2 are measured from left output of the CCD 42-40. The right channel also gives similar noise performance. For DCDS, bandwidth plays a significant role. As can be seen from $eq^n$. 3 and 4 increasing system bandwidth will allow more noise to pass through to the output. But to take as many samples as possible in each sampling period one has to minimize the transition time between two levels which is inversely proportional to the system bandwidth. So it is required to optimize bandwidth and number of samples to get the best noise performance. Table 2 shows that there is a decrease in noise with decrease in system bandwidth until 3MHz beyond which noise is increasing as samples becomes noisy. It is understood that 3MHz is the optimum bandwidth to run the IDSAC system at 500KPPS without deteriorating the signal.